\documentclass[aps,prl,amsfonts,amssymb,twocolumn,showpacs]{revtex4}
\usepackage{mathrsfs}
\usepackage{graphicx,hyperref}

\begin{document}

\title{Nonlinear Coherent Destruction of Tunneling}

\author{Xiaobing Luo}
\affiliation{Institute of Physics, Chinese Academy of Sciences,
Beijing 100080, China}
\author{Qiongtao Xie}
\affiliation{Institute of Physics, Chinese Academy of Sciences,
Beijing 100080, China}
\author{Biao Wu}
\email{bwu@aphy.iphy.ac.cn}
\affiliation{Institute of Physics,
Chinese Academy of Sciences, Beijing 100080, China}

\begin{abstract}
We study theoretically two coupled periodically-curved optical
waveguides with Kerr nonlinearity. We find that the tunneling
between the waveguides can be suppressed in a wide range of
parameters due to nonlinearity. Such suppression of tunneling is
different from the coherent destruction of tunneling in a linear
medium, which occurs only at the isolated degeneracy point of the
quasienergies. We call this novel suppression nonlinear coherent destruction of
tunneling.
This nonlinear phenomenon can be observed readily with current
experimental capability; it may also be observable in a different
physical system, Bose-Einstein condensate.
\end{abstract}

\pacs{42.65.Wi, 42.82.Et, 03.75.Lm, 33.80.Be}
\maketitle

Periodic driving force is an important and effective tool for
coherently controlling quantum tunneling. This has been well
demonstrated with a paradigmatic model, a free  particle in a
double-well potential and driven by a periodic external
field\cite{P.Hanggi}. With appropriately tuned parameters, the
periodic driving force is able not only to enhance
tunneling\cite{Lin}-\cite{Vorobeichik1} but also to completely
suppress it\cite{Grossmann}-\cite{Steinberg}. The latter is rather
surprising and was discovered first by Grossmann {\it et
al}\cite{Grossmann}. It is now known as coherent destruction of
tunneling (CDT)\cite{Grossmann}. When it occurs, a localized wave
packet prepared in one well remains in the same well and does not
tunnel to the other well. In a periodically driven system, there
are Floquet states and associated quasienergies\cite{shirley}. The
CDT is found to occur only at the isolated degeneracy point of the
quasienergies\cite{Grossmann,Grossmann2}.

\begin{figure}[!htb] \center
\includegraphics[width=7.5cm]{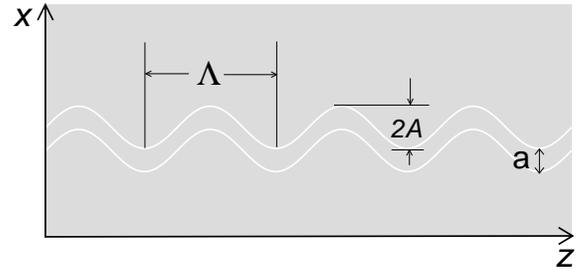}
\caption{Schematic drawing (not to scale) of two periodically
curved optical waveguides placed parallel to each other.}
\label{fig:coupler}
\end{figure}

Recently, this quantum phenomenon of CDT was observed
experimentally with two coupled periodically-curved
waveguides\cite{Longhi2} (see Fig.\ref{fig:coupler}). In this
classical optical system, the Maxwellian wave mimics the quantum
wave while the periodic driving force is achieved by bending the
waveguides periodically. Such a waveguide system is an ideal
laboratory system for demonstrating the coherent control of
quantum tunneling by periodic driving force. For example,
tunneling enhancement has recently also been reported with two
optical waveguides\cite{Vorobeichik}.

In this Letter we consider a similar coupled waveguide system but
with Kerr nonlinearity. With a well-known two-mode approximation,
the system can be described by a two-mode nonlinear model with an
external periodic driving force. This driving is characterized by
two parameters, its frequency $w$ (the inverse of the period of
the curved waveguide) and its strength $S$ (the curving magnitude
of the waveguides) of the driving force. By numerically solving
this two-mode nonlinear model, we find that the suppression of
tunneling between the two coupled waveguides happens for a wide
range of ratio $S/w$. This is in stark contrast to the CDT in
curved linear waveguides that occurs at an isolated point of
$S/w$, where the quasienergies of the system are degenerate. This
extension of tunneling suppression region is caused by
nonlinearity. Therefore, we call it nonlinear coherent destruction
of tunneling (NCDT). We find that the range of ratio $S/w$ for
NCDT increases steeply with nonlinear strength. The Floquet states
and the quasienergies of this nonlinear model are also studied. We
discover that there can be more than two Floquet states and
quasienergies in a certain range of ratio $S/w$. These additional
Floquet states form a triangle in the quasienergy levels. Our
study reveals that these additional Floquet states are closely
related to the NCDT.

The current experimental capability with nonlinear waveguides is
examined. We find that the observation of NCDT is well within the
current experimental ability. Note that the nonlinear two-mode
model that we derived for the waveguides can also be used to
describe the dynamics of a Bose-Einstein condensate in a
double-well potential under a periodic modulation\cite{wang}. This
indicates that NCDT may also be observable with Bose-Einstein
condensates.

In a weakly guiding dielectric structure, the effective
two-dimensional wave equation for light propagation in nonlinear
directional waveguides reads\cite{Micallef}
\begin{eqnarray}
i\frac{\lambda}{2\pi}\frac{\partial \psi}{\partial z}=-\frac{\lambda^2}
{8\pi^2 n_{s}}\frac{\partial^{2} \psi}{\partial
x^2}+V[x-x_{0}(z)]\psi-|\psi|^2\psi.
\label{Eq1}
\end{eqnarray}
where $\lambda$ is the free space wavelength of the light,
$x_0(z)=A\cos(2\pi z/\Lambda)$, and $V(x)\equiv [n_s^2-n^2(x)]/(2
n_s)\simeq n_s-n(x)$, where $n(x)$ and $n_s$ are, respectively,
the effective refractive index profile of the waveguides and the
substrate refractive index. For the coupled waveguides as in
Fig.\ref{fig:coupler}, $n(x)$ thus $V(x)$ have a double-well
structure. The scalar electric field is related to $\psi$ through
$E(x,z,t)=(1/2)(|n_2|n_s\epsilon_0
c_0/2)^{-1/2}[\psi(x,z)\exp(-i\omega t+ikn_sz)+c.c.]$, where $n_2$
is the nonlinear refractive index of the medium, $k=2\pi/\lambda$,
$\omega=kc_0$, and $c_0$ and $\epsilon_0$ are the speed of light
and the dielectric constant in vacuum, respectively. The field
normalization is taken such that $|\psi|^2/|n_2|$ gives the light
intensity $I$ (in $W/m^2$). By means of a Kramers-Henneberger
transformation\cite{kh} $x'=x-x_0(z), z'=z$, and
$\phi(x',z')=\psi(x',z')\exp[-i(2n_s\pi/\lambda)\dot{x}_{0}(z')x'-
i(n_s\pi/\lambda)\int_{0}^{z'}d\xi\dot{x}_{0}^{2}(\xi)]$ (the dot
indicates the derivative with respect to $z'$), Eq.(\ref{Eq1}) is
then transformed to
\begin{eqnarray}
i\frac{\lambda}{2\pi}\frac{\partial \phi}{\partial z'}&=&-\frac{\lambda^2}
{8\pi n_{s}}\frac{\partial^{2} \phi}{\partial
x'^2}+V(x')\phi-|\phi|^2\phi+x'F(z')\phi\nonumber\\&\equiv&
H_0\phi-|\phi|^2\phi+x'F(z')\phi. \label{eq2}
\end{eqnarray}
where  $F(z')=n_s\ddot{x}_0(z')=(4\pi^2 An_s/\Lambda^2)\cos(2\pi
z'/\Lambda)$ is the force induced by waveguide bending. It is
clear that if we view $z$ (or $z'$) as time $t$, the above
equations can be regarded as describing the system of a nonlinear
quantum wave in a double-well potential and under a periodic
modulation.

We assume that the light in each waveguide of the coupler is single moded
and neglect excitation of radiation modes. With a standard two-mode
approximation\cite{Longhi3,khomeriki,Jensen}, we write
\begin{equation}
\phi(x',z')=e^{-\frac{2i\pi}{\lambda}
E_0z'}\Big[c_1(z')u_1(x')+c_2(z')u_2(x')\Big]\,,
\end{equation}
where $u_1$ and $u_2$ are localized waves in two waveguides while
the two coefficients are normalized to one, $|c_1|^2+|c_2|^2=1$.
$E_0$ is defined as $E_0=\int u_{1,2}^*H_0u_{1,2}dx'$. It is
reasonable to assume that the localized wave is a Gaussian,
$u_{1,2}(x')=\sqrt{D}\exp[-(x'\pm a/2)^2/2b^2]$, where $a$ is the
distance between the two waveguides, $b$ is the half-width of each
waveguide, and $D$ is related to the input power of the system
$P(0)$ as $D=n_2P(0)/(\sqrt{\pi}b)$. $P(0)$ has the unit of $W/m$.
The two-mode approximation eventually simplifies Eq.(\ref{eq2}) to
\begin{eqnarray}
i\dot{c_1}&=&\frac{v}{2}c_2-\frac{S}{2}\cos(w z')c_1-\chi|c_1|^{2}
c_1,
\label{twolevel1}\\
i\dot{c_2}&=&\frac{v}{2}c_1+\frac{S}{2}\cos(w z')c_2-\chi|c_2|^{2}
c_2\label{twolevel2},
\end{eqnarray}
where we have set $S=8\pi^3aAn_s/\Lambda^2\lambda$, $v=4\pi(\int
u_1^*H_0u_2dx)/\lambda$, the modulation frequency
$w=2\pi/\Lambda$, and $\chi=\sqrt{2\pi} n_2 P(0)/(\lambda b)$ is
an effective nonlinear coefficient. When $S=0$, Eqs.
(\ref{twolevel1}), (\ref{twolevel2}) will be reduced to the
well-known Jensen equation\cite{Jensen}. Note that $P(0)$ has the
unit of $W/m$ is because the waveguide is two dimensional in our
theoretical model. In experiments, $P(0)$ has the unit of $W$ and
the waveguides are three dimensional. As a result, to relate our
nonlinear parameter to the real experimental parameters, we choose
$\chi=2\pi n_2P(0)/(\lambda\sigma_{\rm eff})$, where $\sigma_{\rm
eff}$ is the effective cross-section of the waveguide, according
to Ref.\cite{Eisenberg}.

\begin{figure}[!htb]
\includegraphics[width=8cm,height=6cm]{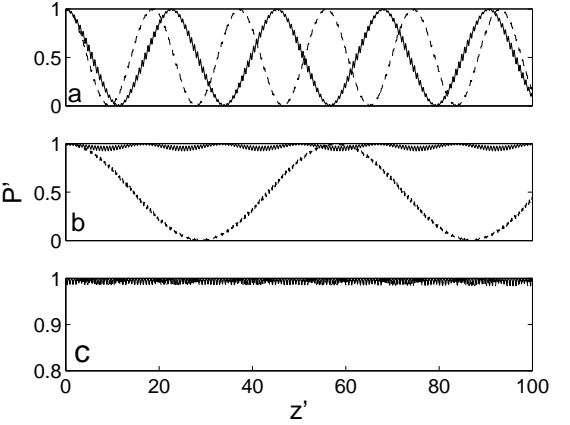}
\caption{The intensity of light in the initially populated waveguide
for the case of $\chi=0$ (dashed lines) and $\chi/v=0.4$(solid
lines) with (a)$S/w=1.8$, (b)$S/w=2.2$, (c)$S/w=2.4$. Distance $z'$
is in units of $1/v$. $w/v=10$.} \label{fig:tunneling}
\end{figure}

To investigate tunneling effect, we solve the two nonlinear
equations (\ref{twolevel1}) and (\ref{twolevel2}) numerically with
the light initially localized in one of the two waveguides. With the
numerical solution, we compute the intensity of the light staying in
the initial well with $P'(z')=|c_1^*(0)c_1(z')+c_2^*(0)c_2(z')|^2$.
Three sets of our results are shown in
Fig.\ref{fig:tunneling}(a,b,c). In the first set for $S/w=1.8$, we
see that $P'(z')$ oscillates between zero and one for both linear
case $\chi=0$ and nonlinear case $\chi/v=0.4$, demonstrating no
suppression of tunneling. In the second set for $S/w=2.2$, we see a
different scenario, the oscillation of $P'(z')$ is limited between
$\sim$0.8 and one for the nonlinear case, showing suppression of
tunneling, while there is no suppression for the linear case. In the
third set for $S/w=2.4$, suppression of tunneling is seen for both
linear and nonlinear cases. Such suppression of tunneling for the
linear case is known as coherent destruction of
tunneling\cite{Grossmann}. These numerical results demonstrate that
nonlinearity can extend the parameter range of the suppression of
tunneling. We call this new phenomenon nonlinear coherent
destruction of tunneling (NCDT).

\begin{figure}[!htb]
\includegraphics[width=7cm]{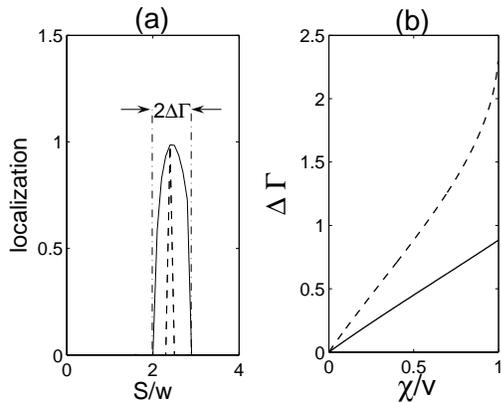}
\caption{(a)Localization as a function of $S/w$. The solid line is
for the nonlinear case $\chi/v=0.4$ and the dashed line is for the
linear case $\chi=0$. $w/v=10$. (b) The width $\Delta\Gamma$ of the
peak in (a) as a function of nonlinearity strength $\chi/v$ (solid
line). The dashed line is for the width of the quasienergy triangle
in Fig.\ref{fig:quasienergies}.} \label{fig:cdtregime}
\end{figure}

The extension of tunneling suppression regime of ratio $S/w$ by
nonlinearity is more clearly demonstrated in
Fig.\ref{fig:cdtregime}(a). In this figure, we have used
localization, which is defined as the minimum value of $P'(z')$,
to measure the suppress of tunneling. When there is large
suppression of tunneling, localization is close to one; when there
is no suppression,localization is zero. As clearly seen in
Fig.\ref{fig:cdtregime}(a), the peak of localization (solid line)
for $\chi/v=0.4$ is much wider than the peak for $\chi/v=0.0$
(dashed line).  In Fig.\ref{fig:cdtregime}(b), we see the width of
localization $\Delta\Gamma$ increases almost linearly with
nonlinearity $\chi$(solid line). Note that, analytically, CDT
occurs only at isolated points. That it has a narrow range in
Fig.\ref{fig:cdtregime}(a) is because the evolution time is finite
in numerical simulation.

\begin{figure}[!htb] \center
\includegraphics[width=8cm]{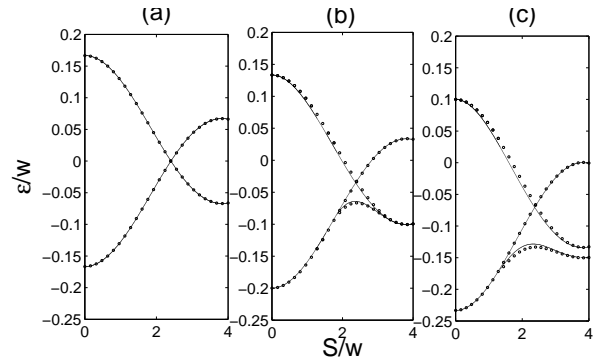}
\caption{Quasienergies at (a) $\chi=0$;(b) $\chi/v=0.4$; (c)
$\chi/v=0.8$. Solid lines are for numerical results obtained with
Eqs.(\ref{twolevel1},\ref{twolevel2}) and circles for the
approximation results for high frequencies with
Eqs.(\ref{highfrequency1},\ref{highfrequency2}). $w/v=3$.
}\label{fig:quasienergies}
\end{figure}

As is well known, the CDT is connected to the degeneracy point of
quasienergies in the system\cite{Grossmann}. Although our system
is nonlinear, one can similarly define its Floquet state and
quasienergy. That is, Eqs.(\ref{twolevel1},\ref{twolevel2}) have
solutions in the form of $\{c_1,c_2\}=e^{-i\varepsilon
z'}\{\tilde{c}_1(z'),\tilde{c}_2(z')\}$, where both $\tilde{c}_1$
and $\tilde{c}_2$ are periodic with period of $\Lambda$. These
Floquet states and corresponding quasienergies $\varepsilon$ can
be found numerically. We first expand the periodic functions
$\tilde{c}_{1,2}$ in terms of Fourier series with a cutoff. After
plugging them into Eqs.(\ref{twolevel1},\ref{twolevel2}), we
obtain a set of nonlinear equations for the Fourier coefficients.
By solving these equations numerically, we obtain the Floquet
states and corresponding quasienergies $\varepsilon$. The results
are plotted in Fig.\ref{fig:quasienergies}, where we witness a
striking difference between the linear and nonlinear cases. As
seen in Fig.\ref{fig:quasienergies}(a), for the linear case, there
are two Floquet states for a given value of $S/w$ and there is
only one isolated degeneracy point. For the nonlinear case, we
notice that there are four Floquet states and three quasienergies
in a certain range of $S/w$ with two of the Floquet states
degenerate. The three quasienergies form a triangle in the
quasienergy levels as seen in Fig.\ref{fig:quasienergies}(b,c).
Our numerical computation shows that the width of the quasienergy
triangle increases with nonlinearity $\chi$ as shown in
Fig.\ref{fig:cdtregime} (dashed line). As this increasing trend is
similar to the localization width $\Delta\Gamma$, this offers us
the first glimpse of link between NCDT and the quasienergies.
Since the right corner of the triangle can be open, we define the
width of the quasienergy triangle as the horizontal distance
between the left corner and the upper corner.

\begin{figure}[!htb]
\includegraphics[width=7cm]{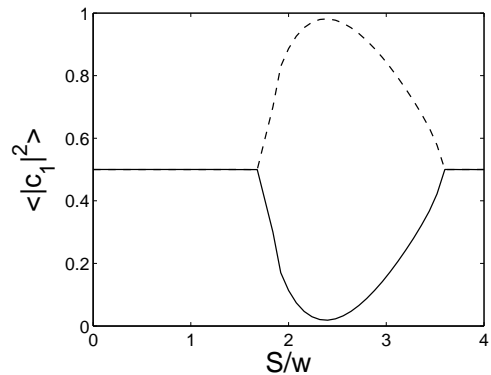}
\caption{Intensity in the first well for every Floquet state in
the lowest quasienergy level at $\chi/v=0.4$, $w/v=3$.}
\label{fig:Floquet}
\end{figure}

A firm link between the NCDT and the triangle structure in the
quasienergies can be established by looking into the Floquet
states. We focus on the Floquet states that correspond to the
lowest quasienergies in Fig.\ref{fig:quasienergies}. To measure
how the Floquet state is localized in one of the two waveguides,
we define
$\langle|c_1|^2\rangle=(\int_{0}^{\Lambda}dz'|c_1|^2)/\Lambda$ for
a given Floquet state $\{c_1,c_2\}$. We have plotted this value
for the lowest Floquet states in Fig.\ref{fig:Floquet}. In this
figure, we see clearly that only the Floquet states on the
quasienergy triangle are localized. This thus demonstrates a clear
link between the quasi-energy triangle and the NCDT. That there
are two lines in Fig.\ref{fig:Floquet} reflects the fact that
there is a two-fold degeneracy for the lowest quasienergies on the
triangle.

The triangular structure in the quasienergy is very similar to the
energy loop discovered within the context of nonlinear
Landau-Zener tunneling\cite{nlz}. In fact, they are mathematically
related. For high frequencies, $w\gg \max\{v,\chi\}$, which is
usually the case for current experiments with optical waveguides,
we take advantage of the transformation
\begin{equation}
\begin{array}{lll}
c_1&=&c_1'\exp[iS\sin(wz')/2w],\cr\cr
c_2&=&c_2'\exp[-iS\sin(wz')/2w].
\end{array}
\label{trans}
\end{equation}
After averaging out the high frequency terms\cite{wang}, we
find a non-driving nonlinear model,
\begin{eqnarray}
i\dot{c_1'}&=&\frac{v}{2}J_0(S/w)c_2'-\chi|c_1'|^{2} c_1',
\label{highfrequency1}\\
i\dot{c_2'}&=&\frac{v}{2}J_0(S/w)c_1'-\chi|c_2'|^{2}
c_2'\label{highfrequency2},
\end{eqnarray}
where $J_0$ is the zeroth-order Bessel function.  It is clear from
the transformation in Eq.(\ref{trans}) that the eigenstates of the
above time-independent nonlinear equations correspond to the
Floquet states of Eqs.(\ref{twolevel1},\ref{twolevel2}). We have
computed the eigenstates of
Eqs.(\ref{highfrequency1},\ref{highfrequency2}) and the
corresponding eigenenergies, which are plotted as circles in
Fig.\ref{fig:quasienergies}. The consistency with the previous
results is obvious. As is known in Ref.\cite{nlz}, the above
nonlinear model admits additional eigenstates when
$\chi>J_0(S/w)v$. Therefore, this can be regarded as the condition
for the extra Floquet states to appear for the driving nonlinear
model Eqs.(\ref{twolevel1},\ref{twolevel2}) at high frequencies.


So far, we have focused on self-focusing materials. Our approach
and results will be very similar if one considers instead
self-defocusing materials, for which the sign before the nonlinear
term in Eq.(\ref{Eq1}) should be plus. Nonlinear coherent
destruction of tunneling still occurs and the triangular structure
also appears in the quasienergy levels but its direction is
reversed as compared to the self-focusing case.

At present the nonlinear waveguides are readily available in
labs\cite{Eisenberg,Al-hemyari,Friberg}. We take the experimental
parameters in Ref.\cite{Al-hemyari} to estimate our theoretical
values in Eqs.(\ref{twolevel1},\ref{twolevel2}). The wavelength of
the laser light is $\lambda=1.55\mu$m, the effective
cross-sectional area of the waveguide is $\sigma_{\rm
eff}=12\mu$m$^2$, the nonlinear index $n_2=1.2\times 10^{-13}{\rm
cm}^2/W$, and the shortest length for the light transfer from one
waveguide to the other waveguide in the weak nonlinearity limit is
$L_c\approx 2$cm. With the power input in the waveguides $P(0)\sim
100W$, we have
\begin{equation}
\frac{\chi}{v}=\frac{2\pi n_2P(0)L_c}{\pi\lambda\sigma_{\rm
eff}}\approx 2\,.
\end{equation}
This shows that strong nonlinear waveguides are available at
optical labs and nonlinear coherent destruction of tunneling can
be visualized in an optical experiment similar to the one in
Ref.\cite{Longhi2}. We also want to mention briefly that NCDT may
be applied to improve optical switching
devices\cite{Al-hemyari,Friberg}. The details will be discussed
elsewhere.

In conclusion, we have studied the light propagation in a
nonlinear periodically-curved waveguide directional coupler. We
have found a new type of suppression of tunneling in this system,
which is induced by nonlinearity and has no linear counterpart. We
call it nonlinear coherent destruction of tunneling (NCDT) in
analogy to a similar but different phenomenon in linear driving
systems, coherent destruction of tunneling. The NCDT occurs for an
extended range of ratio $S/w$, where $S$ is the strength of the
driving and $w$ is its frequency. We have also found that the NCDT
is closely related to a triangular structure appeared in the
quasienergy levels of the nonlinear system. We have also pointed
out that observation of the novel nonlinear phenomenon is well
within the capacity of current experiments.

This work is supported by NSF of China (10504040), the 973 project
of China(2005CB724500,2006CB921400), and  the ``BaiRen" program of
Chinese Academy of Sciences.


\begin{thebibliography}{99}
\bibitem{P.Hanggi}  M. Grifoni, and P.H\"{a}nggi, Phys. Rep. \textbf{304},
229(1998).
\bibitem{Lin} W. A. Lin and L. E. Ballentine, Phys. Rev. Lett. \textbf{65},
2927 (1990).
\bibitem{Peres} A. Peres, Phys. Rev. Lett. \textbf{67}, 158 (1991).
%
\bibitem{Vorobeichik1} I. Vorobeichik and N. Moiseyev, Phys. Rev. A, \textbf{59}, 2511 (1999).
%
\bibitem{Grossmann} F. Grossmann, T. Dittrich, P. Jung, and P. Hanggi, Phys. Rev.
Lett. \textbf{67}, 516 (1991); Z. Phys. B \textbf{84}, 315 (1991).
%
\bibitem{Grossmann2} F. Grossmann and P. Hanggi, Europhys. Lett. \textbf{18}, 571 (1992).
%
\bibitem{Bavli} R. Bavli and H. Metiu, Phys. Rev. Lett. \textbf{69}, 1986 (1992).
%
\bibitem{Steinberg} M. Steinberg and U. Peskin, J. Appl. Phys. \textbf{85}, 270 (1999).
%
\bibitem{shirley}J. H. Shirley, Phys. Rev. {\bf 138}, B979 (1965).
%
\bibitem{Longhi2} G. Della Valle, M. Ornigotti, E. Cianci, V. Foglietti, P. Laporta,
and S. Longhi. e-print arXiv: quant-ph/0701121.
%
\bibitem{Vorobeichik} I. Vorobeichik, E. Narevicius, G. Rosenblum, M. Orenstein,
and N. Moiseyev, Phys. Rev. Lett. \textbf{90}, 176806 (2003).



\bibitem{wang} Guan-Fang Wang, Li-Bin Fu and Jie Liu, Phys. Rev. A \textbf{73}, 013619(2006).
\bibitem{Micallef} R. W. Micallef, Y. S. Kivshar, J. D. Love, D. Burak, and R. Binder, Opt. Quantum
Electron. \textbf{30}, 751 (1998).
\bibitem{kh}W.C. Henneberger, Phys. Rev. Lett. \textbf{21}, 838 (1968).
\bibitem{Longhi3} S. Longhi, Phys. Rev. A \textbf{71}, 065801 (2005).

\bibitem{khomeriki} R. Khomeriki, J. Leon, and S. Ruffo, Phys. Rev. Lett. \textbf{97}, 143902 (2006).
\bibitem{Jensen}  S.M. Jensen, IEEE J. Quantum Electron. QE-\textbf{18}, 1580
(1982).

\bibitem{Eisenberg} H. S. Eisenberg, Y. Silberberg, R. Morandotti, A. R. Boyd, and J. S. Aitchison, Phys. Rev.
Lett. \textbf{81}, 3383 (1998).
%
\bibitem{Al-hemyari} K. Al-hemyari, A. Villeneuve, J.U. Kang, J.S. Aitchison, C.N.
Ironside, G.I. Stegeman, Appl. Phys. Lett. \textbf{63}, 3562 (1993).
%
\bibitem{Friberg} S. R. Friberg, Y. Silberberg, M.K. Oliver, M.J. Andrejco, M.A. Saifi, P.W.
Smith, Appl. Phys. Lett. \textbf{51}, 135 (1987).
%
\bibitem{nlz}B. Wu and Q. Niu, Phys. Rev. A \textbf{61}, 023402(2000).



\end{thebibliography}
\end{document}